\documentclass[letterpaper,twocolumn,10pt]{article}
\PassOptionsToPackage{hyphens}{url}\usepackage{hyperref}
\usepackage{usenix2019_v3,epsfig,endnotes}
\usepackage{xcolor}

\usepackage{xspace}
\usepackage{relsize}
\newcommand{\code}[1]{\texttt{#1}}
\def\ifmonospace{\ifdim\fontdimen3\font=0pt }
\def\C++{%
\ifmonospace%
    \C++%
\else%
    C\kern-.1167em\raise.30ex\hbox{\smaller{++}}%
\fi%
\spacefactor1000 }

\usepackage{xspace}
\newcommand{\toolname}{DMON\xspace}
\newcommand{\mastermonitor}{L-MON\xspace}
\newcommand{\slavemonitor}{F-MON\xspace}
\newcommand{\connector}{RC-COM\xspace}
\newcommand{\daemon}{NVX-DAEMON\xspace}
\newcommand{\appstarter}{APP-STARTER\xspace}

\usepackage{makecell}
\usepackage{multirow}
\usepackage{subfigure}
\usepackage{tikz}
\usepackage{pgfplots}
\usepackage{pgfplotstable}

\newcommand*\circledw[1]{\tikz[baseline=(char.base)]{
		\node[shape=circle,draw,inner sep=.75pt, line width=.7pt] (char) {\scriptsize #1};}}

\newcommand{\numofsupsyscalls}{93}

\begin{document}

\date{}

\title{\Large \bf \toolname: A Distributed Heterogeneous N-Variant System}

\author{
{\rm Alexios Voulimeneas}\\
University of California, Irvine
\and
{\rm Dokyung Song}\\
University of California, Irvine
\and
{\rm Fabian Parzefall}\\
University of California, Irvine
\and
{\rm Yeoul Na}\\
University of California, Irvine
\and
{\rm Per Larsen}\\
University of California, Irvine
\and
{\rm Michael Franz}\\
University of California, Irvine
\and
{\rm Stijn Volckaert}\\
imec-DistriNet, KU Leuven
} 

\maketitle


\subsection*{Abstract}
N-Variant Execution (NVX) systems utilize software diversity techniques for
enhancing software security.  The general idea is to run multiple
\emph{different} variants of the same program alongside each other while
monitoring their run-time behavior.  If the internal disparity between the
running variants causes observable differences in response to malicious inputs,
the monitor can detect such divergences in execution and then raise an alert
and/or terminate execution.

Existing NVX systems execute multiple, artificially diversified program variants
on a single host.
This paper presents a novel, distributed NVX design that executes program
variants across multiple heterogeneous host computers; our prototype
implementation combines an x86-64 host with an ARMv8 host.  Our approach greatly
increases the level of ``internal different-ness'' between the simultaneously
running variants that can be supported, encompassing different instruction sets,
endianness, calling conventions, system call interfaces, and potentially also
differences in hardware security features.

A major challenge to building such a heterogeneous distributed NVX system is
performance.  We present solutions to some of the main performance challenges.
We evaluate our prototype system implementing these ideas to show that it can
provide reasonable performance on a wide range of realistic workloads.

\section{Introduction}

Memory errors have been a continuous source of software vulnerabilities for C
and \C++ programs.
Attackers and defenders are engaged in an arms race in which the former keep
developing increasingly sophisticated defenses while their adversaries keep
crafting novel exploits that bypass these defenses~\cite{szekeres.etal+13}. At
present, adversaries rely on intimate knowledge of the target environment (such as details about the victim
application, the target operating system, the instruction set
architecture of the host, and runtime parameters such as memory addresses) to
mount code-reuse~\cite{Shacham2007, Shacham10, snow.etal+13} or data-oriented
attacks~\cite{dataonly, hu2016data, hu2015automatic} that allow them to take
control of the target and/or leak its sensitive data.  While memory safety
techniques can provide strong protections against these threats, many of these
techniques have not seen widespread deployment due to
performance~\cite{nagarakatte2009softbound, nagarakatte2010cets} and
compatibility problems~\cite{song2019sanitizing}.

Instead, defenders resort to mitigation techniques that have a more
reasonable performance impact, e.g., control-flow integrity (CFI)
techniques~\cite{abadi2005control, burow2017control}, automated software
diversity techniques~\cite{larsen2014sok}, or a combination thereof. Both of
these classes of defenses have a history of known weaknesses. CFI techniques
often leave sufficient leeway to mount attacks on all but the most trivial
applications~\cite{goktas2014out, goktas2014size, ROP-CFI:Davi,
carlini2015controlflowbending, conti2015losing, evans2015control,
vanderveen2017dynamics, farkhani2018typebasedcfibypass}, and software diversity
techniques have been bypassed using brute-forcing and information leakage attacks,
including attacks enabled by micro-architectural side channels~\cite{Shacham04, snow.etal+13,
bittau2014hacking, conti2015losing, gras2017aslr, goktacs2016undermining,
oikonomopoulos2016poking, evans.etal+15, gawlik2016enabling, barresi2015cain,
hund2013practical, gruss2016prefetch, jang2016breaking, evtyushkin2016jump,
lee2017inferring}.

N-Variant eXecution (NVX) systems amplify the effectiveness of software
diversity techniques and increase resilience~\cite{berger2006diehard,
  bruschi2007diversified, cox2006n, salamat2009orchestra, volckaert2012ghumvee,
  maurer2012tachyon, hosek2013safe, kim2015dual, hosek2015varan, kwon2016ldx,
  koning2016secure, volckaert2016cloning, volckaert2016secure, xu2017bunshin,
  lu2018stopping, osterlund2019kmvx}. An NVX system runs multiple diversified
variants of the same program in parallel on the same inputs while monitoring the
variants' behavior for divergences. With the right selection of diversity
techniques, NVX can make successful exploitation substantially harder (and, in
some cases, even provably impossible) as it forces adversaries to simultaneously
compromise multiple program variants without causing observable changes in their
behavior.

\begin{figure}[t!]
	\centering
	\includegraphics[width=.8\columnwidth]{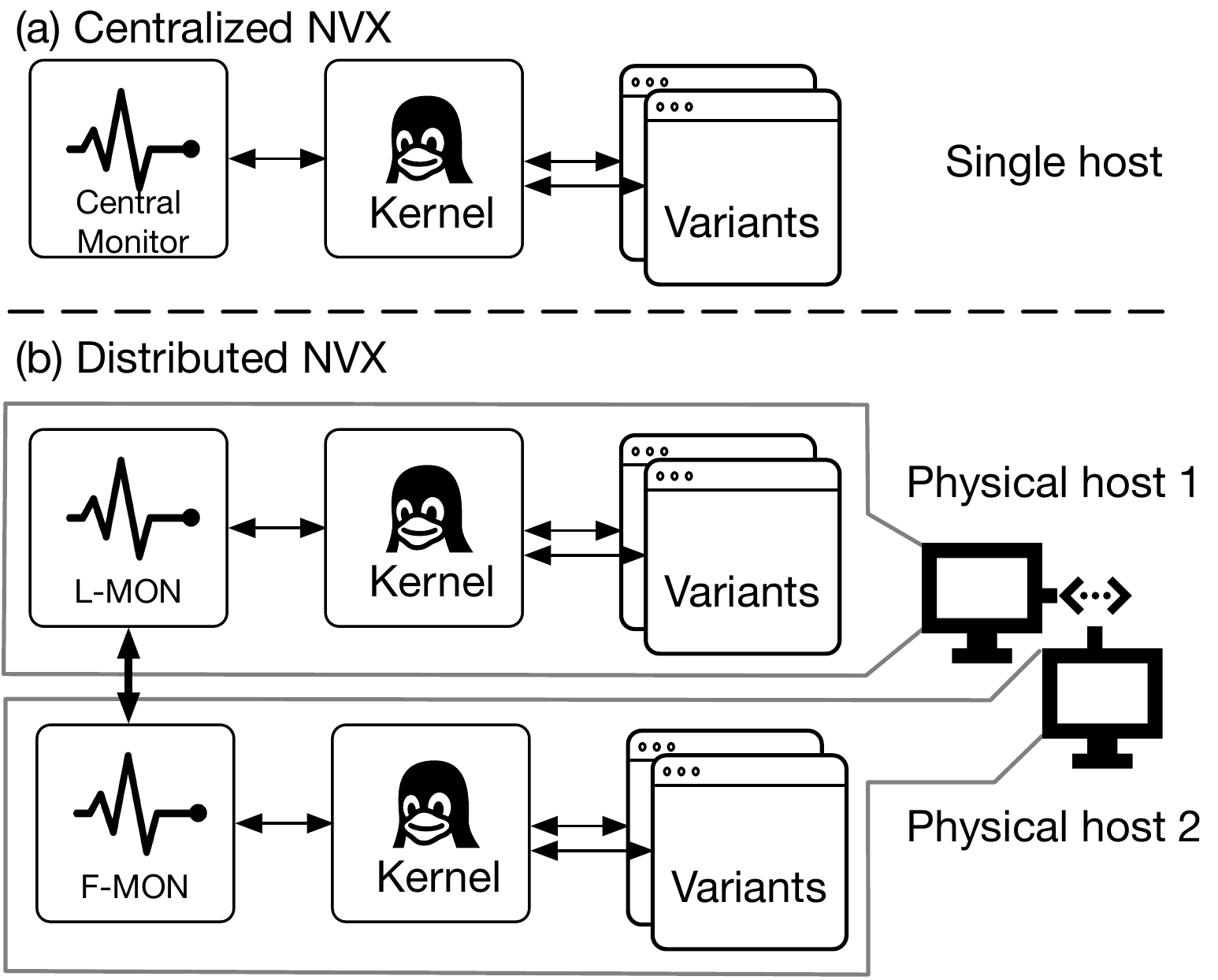}
	\vspace{-.4cm}
	\caption{Two NVX designs running two variants. 
	}
	\label{fig:NVX-overview}
	\vspace{-1.50em}
\end{figure}

Existing NVX systems have been particularly effective at stopping attacks that
rely on knowledge of the target's absolute virtual address space layout (i.e.,
code-reuse exploits whose payloads include absolute pointer
values)~\cite{cox2006n,bruschi2007diversified,volckaert2016secure}, as well as
attacks that attempt to acquire that knowledge (i.e., information leakage
attacks)~\cite{lu2018stopping}. However, these systems are not resilient to
Position-Independent Return-Oriented Programming (PIROP)
attacks~\cite{goktas2018pirop} and certain Data-Oriented Programming (DOP)
attacks~\cite{hu2016data}, which build on knowledge of the program's internal
geometry (e.g., relative data/instruction layouts) and/or data representation.
The main reason is that in previous NVX systems all the variants run on the same
machine as shown in Figure~\ref{fig:NVX-overview}(a). Thus, the amount of
diversity that such systems can achieve is limited to what a single platform can
offer. On the other hand, binaries targeted at multiple different platforms have
an inherent diversity that comes naturally from differences in calling
conventions, instruction set architectures, endianness, system call interfaces
and available hardware features.

In this paper, we present \toolname{}, an NVX system that allows us to leverage
the diversity that naturally exists across different platforms, thereby
increasing resilience to memory exploits. \toolname{} compiles and runs each
program variant on its own dedicated machine and monitors divergent behavior
between these distributed variants by cross-checking them at the system call
boundary via a network. Figure~\ref{fig:NVX-overview}(b) illustrates our
design. To bypass \toolname{}, adversaries would need to develop exploits that
work simultaneously against the two (or more) different ISAs and ABIs that the
program variants are compiled for. \toolname{} runs on commodity hardware with
regular multi-core CPUs. As our evaluation shows, \toolname{}'s overhead can be
further reduced by optionally adding specialized network interface cards with
Remote Direct Memory Access (RDMA) support.

Our contributions are as follows: 

\begin{itemize}
\item We present \toolname{}, the first system that combines ISA and ABI
  heterogeneity with N-Variant Execution. \toolname{} distributes the execution
  of a set of variants over a heterogeneous set of physical machines. \toolname{}
  provides a natural resilience against memory exploits as it forces
  adversaries to develop exploits that work simultaneously against multiple
  ISAs, ABIs, system call interfaces, and available hardware features. 

\item We study and identify several performance bottlenecks and
  cross-checking issues that are unique to the distributed and heterogeneous
  monitoring setting and present strategies to alleviate these issues.

\item We evaluate \toolname{}'s security on several realistic server
  applications and show that \toolname{} makes successful code-reuse and data-only
  attacks substantially more difficult.

  

\item We evaluate the performance of \toolname{} on a wide set of
  microbenchmarks and server applications, conduct a thorough security
  analysis on several sets of ISA/ABI-heterogeneous program variants, and conclude
  that \toolname{} offers strong protections at reasonable cost.
\end{itemize}

\section{Background}

Researchers in the information security~\cite{berger2006diehard,
  bruschi2007diversified, cox2006n, salamat2009orchestra, volckaert2012ghumvee,
  koning2016secure, volckaert2016cloning, volckaert2016secure, xu2017bunshin,
  lu2018stopping, osterlund2019kmvx} and software reliability
communities~\cite{maurer2012tachyon, hosek2013safe, kim2015dual, hosek2015varan,
  kwon2016ldx, pina2019mvdsua} have presented over a dozen different NVX systems
since 2006.  Although these systems serve a variety of different purposes, they
do have some essential similarities. First, all systems have the same high-level
architecture; two or more software variants execute simultaneously on the same
physical machine, while a monitoring component (on that same machine) compares
the variants' overall behavior, provides them with identical inputs, and
demultiplexes their outputs. Most monitors perform these tasks by forcing the
variants to execute in lock-step at the granularity of system calls. This means
that the variants will be suspended at every system call entry and exit, and
they will not be allowed to proceed until the monitor has cross-checked (i.e.,
compared across variants) the system call numbers and arguments.

Second, all existing NVX systems cross-check behavior and replicate I/O by
intercepting the variants' system calls. Most early systems used a dedicated
monitoring process that attaches to the variants and intercepts their system
calls using the \texttt{ptrace} API~\cite{bruschi2007diversified,
  salamat2009orchestra, volckaert2012ghumvee, maurer2012tachyon,
  hosek2013safe}. To avoid the high run-time performance overhead
incurred by context switching between a variant process and the monitor process,
several teams explored alternative designs that use binary
rewriting~\cite{hosek2015varan}, virtualization
features~\cite{koning2016secure}, or kernel modules~\cite{cox2006n,
  volckaert2016secure, xu2017bunshin, lu2018stopping} to intercept and
cross-check system calls more efficiently, within the variants' processes and
address spaces.

\vspace{-1em}
\subsection{System Calls and I/O Replication}

Modern operating systems isolate processes from one another by providing them
with a distinct virtual address space and limited privileges. Processes must
use system calls to interact with other processes, the host system, and its
hardware resources. While the exact implementation of this system call
interface depends on the instruction set and operating system, it generally
works as follows. First, the program loads the system call number and the
system call arguments into the appropriate processor registers or into the
stack. Then, the program executes the \emph{system call} instruction defined by
the processor's instruction set. Next, the CPU raises the current privilege
level and transfers control to the kernel's system call entrypoint. The kernel
then reads the system call number and arguments and invokes the appropriate
system service. Finally, the kernel executes the \emph{system call return}
instruction to lower the privilege level and to transfer control back to the
program.

NVX systems monitor behavior and replicate I/O at the system call
interface. This design lets the system monitor all behavior that can
potentially affect the integrity of the OS or other processes, as well as all
communication between the variants and external entities\footnote{Communication
  via shared memory is not visible at the system call interface, so most NVX
  systems prevent the variants from mapping shared memory regions.}. The
monitoring and replication must be transparent to the program variants and to
the end-user. In other words, neither the variants, nor any external observer
should be able to notice any differences (apart from timing) between native
execution of a single variant and NVX of multiple variants. To provide this
guarantee, our system designates one variant as the leader, while the others
become followers.  Whenever the variants attempt an I/O operation, our system
ensures that only the leader variant actually completes the operation, while the
followers skip the operation and wait until they receive the I/O results from
the monitor.


\subsection{ISA-Heterogeneity}



%
An underlying assumption of NVX is that the program variants will behave
identically if i) they are built from the same source code, and ii) they receive
equivalent benign inputs. This assumption no longer holds in our setting, where
we run variants on processors with different ISAs. Differences in the
endianness, register, and pointer width --- and even the available system calls
could lead to observable (yet benign) differences in the variants' behavior,
which would all cause false alarms in a traditional NVX system. We encountered
many such differences and designed \toolname{} so it can tolerate any expected
divergences that stem from the heterogeneous-ISA setting.

\subsection{ABI-Heterogeneity}

Aside from obvious differences such as different instruction opcodes and
encoding, endianness, and register/pointer width, ISA-heterogeneous variants
often also differ in more subtle ways because the OS maintainers impose a set of
conventions for all binary programs compiled for the target ISA.  The
Application Binary Interface (ABI) documents rules such as sizes of primitive
data types, how structs are packed, padded, and aligned, how function callers
can pass arguments to their callees, etc. Many of these conventions also affect
the program behavior as observed from the system call interface, and we
therefore had to carefully design \toolname{} so it takes the ABIs into account
when comparing variant behavior.


\section{Threat Model}



Throughout the rest of the paper, we will make the following assumptions about
the host system and the attacker. Our assumptions are consistent with related
work in this area~\cite{volckaert2016secure}.

\paragraph*{\textbf{Host defenses}} We assume that the standard set of
migitations are in place on any of the physical machines \toolname{} and the
variants run on. Specifically, we assume that Data Execution Prevention (DEP) is
used, and that memory pages are therefore never writable and executable at the
same time. DEP therefore rules out code-injection attacks. Likewise, we assume
that all of the host systems have Address Space Layout Randomization (ASLR)
enabled. ASLR randomizes the base addresses of the main program executable and
shared libraries, as well as the heap, stack, and any other mapped memory
regions. Note that, even with ASLR in place, the NVX system can still override
the base addresses of any region mapped into a variant's address
space~\cite{volckaert2016cloning,lu2018stopping}.

\paragraph*{\textbf{Known and vulnerable target}} We assume that the protected
application is known to the attacker, and that the attacker either has direct
access to the variant binaries, or that the attacker can reproduce exact
replicas of any of the target binaries for offline analysis. We further assume
that the protected application has an arbitrary memory read/write vulnerability
that the attacker knows how to trigger.

\paragraph*{\textbf{Remote attacker}} We assume that the attacker does not have
direct access to any of the physical machines \toolname{} (or the variants) run
on. The attacker can only communicate with the protected application via a
remote communication channel such as a network socket. Because the attacker is
remote, we also assume that any run-time secrets embedded into the variants
(e.g. randomized base addresses) are not known a priori. 

\section{\toolname{} Design}
\label{sec:design}

\toolname{} orchestrates and supervises the execution of a set of diversified
program variants running natively on machines that differ in their instruction
set architecture. Like most other NVX systems, \toolname{} uses a
leader/follower-model for I/O replication. The designated leader variant is the
only variant allowed to perform externally observable I/O operations such as
sending or receiving data from a network socket. \toolname{} forces follower
variants to skip these I/O operations and instead provides them with the
leader's I/O results, thus emulating the original operation unbeknownst to the
follower.

Similar to other security-focused NVX systems such as
ReMon~\cite{volckaert2016secure} and MvArmor~\cite{koning2016secure},
\toolname{} executes all security sensitive system calls in lock step. Whenever
the variants attempt to execute a sensitive system call, \toolname{} ensures
that the variants can neither enter the system call routine, nor exit from it
until \toolname{} has ensured that all variants have reached equivalent states.
We distinguish between the following components of a running \toolname\ system:

\begin{enumerate}
\item \textbf{Leader Variant}  Only the designated leader variant is allowed to
  perform externally observable I/O. As in any other NVX system, \toolname{}
  requires that there is exactly one leader variant. The leader designation is
  fixed. Leader variants cannot become followers and vice versa.

\item \textbf{Follower Variants} Follower variants skip externally observable
  I/O operations and use the leader's I/O results instead. \toolname{} supports
  any number of such variants.

\item \textbf{Monitors} The monitors are responsible for starting the variants,
  supervising their execution, exchanging system call metadata (system call
  numbers, arguments, and results), performing security checks, and enforcing
  lock-step execution. \toolname{} uses two types of monitors: the (single)
  \mastermonitor{} monitor supervises the leader variant, while every follower
  variant is supervised by its own \slavemonitor{} monitor.

\item \textbf{\connector{}} A reliable communication component used to exchange
  system call metadata between the monitors. By separating the communication
  logic into its own abstraction layer, we have enabled the monitors to
  communicate over a variety of communication channels (e.g., loopback
  interfaces vs network cards, different network protocols, etc.).
\end{enumerate}

\begin{figure}[t]
	\centering
	\includegraphics[width=\columnwidth]{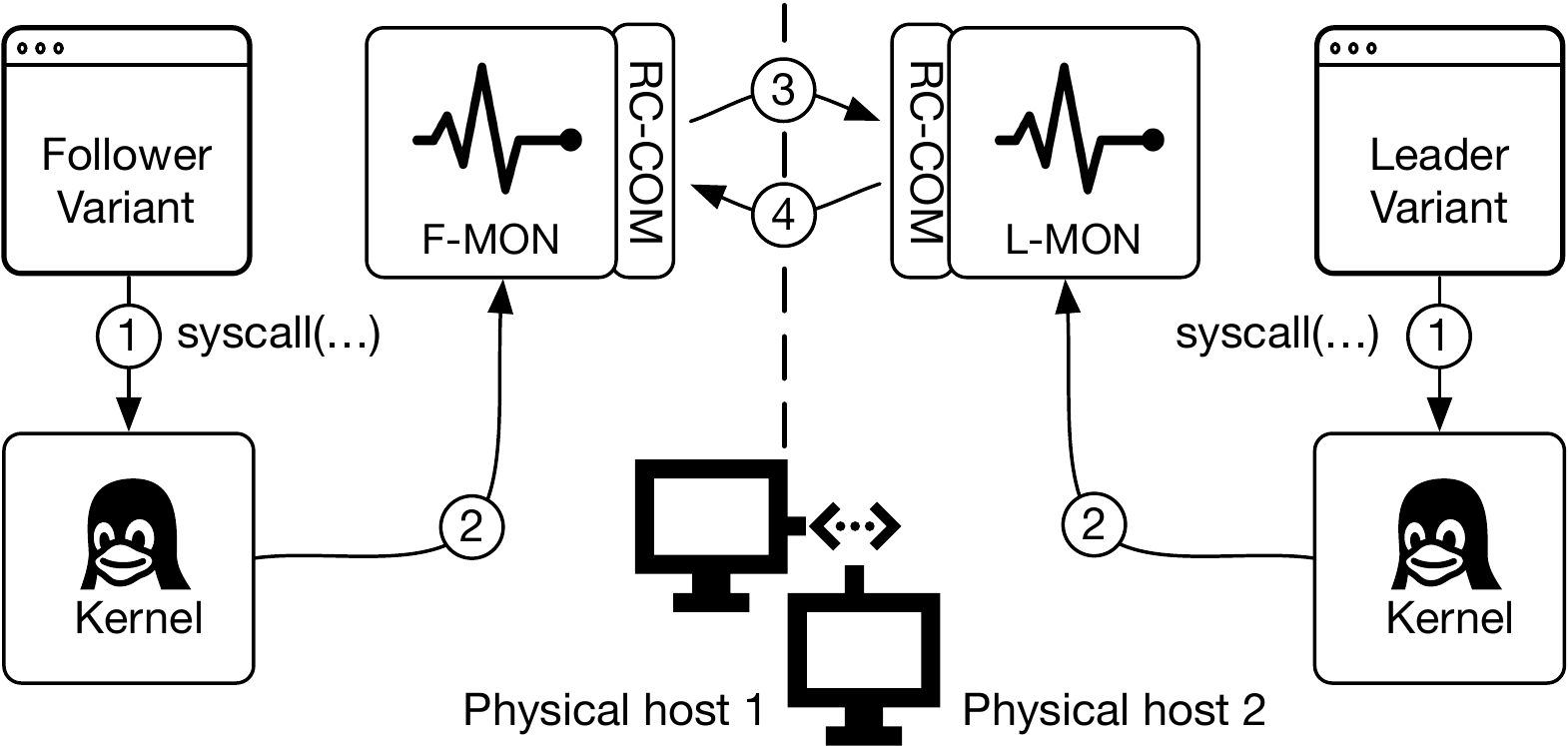}
	\vspace{-.8cm}
	\caption{\toolname's basic components and interactions.}
	\label{fig:NVX-design}
	\vspace{-1.50em}
\end{figure}

These components interact whenever the variants execute system calls, as shown
in Figure~\ref{fig:NVX-design}. Whenever a leader or follower variant attempts
to enter or exit from a system call (\circledw{1}), the corresponding
\mastermonitor{} or \slavemonitor{} interrupts and suspends the variant, reads
the call number of the interrupted system call, and invokes a specialized
handler routine within the monitor process (\circledw{2}), which implements the
cross-checking and replication logic for that system call.

The monitors use cross-checking handlers when they interrupt variants upon
entering a system call. In \slavemonitor{}, the cross-checking handler gathers
information about the variant's state, sends this information to
\mastermonitor{} (\circledw{3}), and waits for \mastermonitor{} to confirm that
the follower variant is in a state equivalent to the leader variant
(\circledw{4}). In \mastermonitor{}, the cross-checking handler waits for
incoming state information from \slavemonitor{}, compares that state information
with the leader variant's state, and informs \slavemonitor{} about the results
of the comparison.

The state information consists of system call numbers and arguments, with the
latter often consisting of one or more pointers to complex data structures
(e.g., I/O vectors). The cross-checking handlers serialize these corresponding
data structures and append the serialized data to the state information, thereby
allowing \mastermonitor{} to check the variant states for deep equivalence (two
data structures are deeply equivalent when the raw data they contain is
identical, even though the data or the data structures may be stored at
different addresses).  If the variant states do not match, \toolname{} will
interpret that as a sign of potential compromise, and it will abort execution to
protect the host system.

Naive cross-checking of these variant states could trigger false alarms for
divergent behavior because the system call interfaces, calling conventions, data
representation, etc. may differ across platforms. \toolname{}'s cross-checking
handler reconciles such differences to avoid alarms for the expected (and benign)
divergences (see Section~\ref{sec:divergences}). For example, the same system
call may have different system call numbers in different platforms. To correctly
handle this, \toolname{} keeps a mapping between these syscall numbers. The
cross-checking handler consults this mapping to recognize equivalent
system calls between variants running on different platforms.

If the states do match, the cross-checking handler allows the leader variant to
proceed and to enter the kernel-space system call routine. The follower variants
can also proceed, but may (optionally) see their system call number replaced by
that of the \texttt{sys\_getpid} routine in case they attempt to perform an
externally observable I/O operation. This mechanism for skipping system calls
was also used in prior work~\cite{salamat2009orchestra}.

The monitors use replication handlers when they interrupt variants that return
from a system call. Replication handlers for I/O system calls broadcast the
system call results from the leader variant to the followers. Replication
handlers for other system calls are generally no-ops.

\subsection{Monitor Design}

Prior work often used a central monitor process which simultaneously supervised
all of the variants~\cite{salamat2009orchestra, volckaert2012ghumvee,
bruschi2007diversified}. Subsequent research showed that this centralized model
was overly focused on simplicity and security at the expense of performance,
and suggested various designs in which each variant was supervised by a
dedicated monitor
instance~\cite{hosek2015varan,volckaert2016secure,koning2016secure,xu2017bunshin,
lu2018stopping, osterlund2019kmvx}. This dedicated monitor
instance could be loaded directly into the variants' address spaces, thereby
sacrificing the isolation between the variants and the monitor for greatly
reduced variant-monitor communication overhead.

\toolname{} combines elements of both designs. Since we run ISA-heterogeneous
variants on different physical machines, we cannot use a single (central)
monitor that attaches locally to all variants. Instead, we use a dedicated
monitor for each variant and run the monitor on the same machine as the variant
it supervises. Our design does, however, enforce strict isolation between the
variant and its monitor by running the monitor as a separate process that
attaches to the variant using the \texttt{ptrace} API.

Each monitor includes an \connector{}, which exposes our inter-monitor
communication API. By separating the low-level communication logic from the
monitor, we were able to implement and compare various communication mechanisms.

\subsection{Inter-Monitor Communication}
\label{sec:communication}

\slavemonitor{} and \mastermonitor{} communicate whenever the variants execute a
system call. This exchange may include system call numbers, serialized system
call arguments, system call results, or instructions on how to proceed from a
system call entry point (see Section~\ref{sec:design}). In many cases,
particularly when the system call being executed is deemed security-sensitive,
communication must happen synchronously. For instance, \mastermonitor{} cannot
allow the leader variant to proceed past a system call entry point until all
instances of \slavemonitor{} have serialized the state of their corresponding
variant, and until they have sent this state to \mastermonitor{}.
\slavemonitor{} needs to wait even longer as it cannot allow the follower
variants to proceed until \mastermonitor{} has compared the variant states and
it has received \mastermonitor{}'s confirmation that the states match.

For good performance, \toolname{} therefore requires a reliable inter-monitor
communication channel with minimal latency and high bandwidth. There are many
ways to realize such a channel. We experimented with various designs of this
communication channel and implemented them in our \connector{}, which
exposes the inter-monitor communication API to our monitors. 

\paragraph*{\textbf{Network Protocol Choice.}}
The most obvious protocol that meets our reliability demands is TCP, which we
used as the basis for our first implementation of \connector{}. However, even
with extensive tweaking, our TCP-based implementation had poor throughput and
high latency. As an alternative, we therefore used ENet, a lightweight UDP-based
protocol that also offers reliable in-order and error-free data
transfer~\cite{enet}. Our performance evaluation confirms that ENet is more
efficient than the TCP-based implementation that uses the standard TCP/IP stack.

\paragraph*{\textbf{User-Space Networking.}} Besides the networking hardware,
the operating system also impacts the communication bandwidth and latency. When
a network adapter receives a packet, for example, the OS first stores the packet
in a kernel-space buffer, before copying it into the receiving application's
memory and transferring control to the application. These extra copy operations
can be avoided with techniques such as Remote Direct Memory Access (RDMA). RDMA
allows two communicating peers to read or write directly from or to the other
peer's application memory, thus bypassing the kernel's networking stack. We
implemented an RDMA-based version of our \connector{} using Mellanox ConnectX
100 gigabit ethernet
interfaces\footnote{\url{https://store.mellanox.com/products/mellanox-mcx515a-ccat-connectx-5-en-network-interface-card-100gbe-single-port-qsfp28-pcie3-0-x16-tall-bracket-rohs-r6.html}}
and the Mellanox Messaging Accelerator user-space networking
library\footnote{\url{https://github.com/Mellanox/libvma/}}.

\subsection{Optimizations}
\label{sec:optimizations}

To improve \toolname{}'s performance even further, we implemented several
optimizations that can reduce the number of the data packets exchanged by our
monitors.

\paragraph*{\textbf{Permissive Filesystem Access.}} Traditional NVX systems
enforce replication for \emph{all} I/O operations, regardless of the type of I/O
resource being accessed. The system allows one variant to effectively perform
the operation and it then replicates the results to the other variants.  Even
though this replication mechanism seamlessly provides identical inputs to all
variants, it is not always necessary in \toolname{'s} case. Specifically, there
is no need to replicate read accesses to files that were identical on all
physical machines when \toolname{} started, as long as the files have not been
modified while \toolname{} was running. We refer to such files as \emph{static
  files} and designed \toolname{} such that it can identify them. We also
designed the cross-checking handlers for read-only operations such as
\texttt{sys\_read}, \texttt{sys\_readv}, \texttt{sys\_pread}, and
\texttt{sys\_fstat} so that all variants may (optionally) read static files
directly from their local file system, thus bypassing the I/O replication
mechanism.

To support this optimization, \toolname{} requires that the application's root
directory has the same path name on all machines. \toolname{} further assumes
that the all files except executables and shared libraries in the application's
root- and subdirectories are identical when the system starts.

\paragraph*{\textbf{Asynchronous Cross-Checking.}} Our basic approach described
in Section~\ref{sec:design} adds considerable overhead to every system call
invocation as every cross-check happens synchronously and requires at least two
network round-trips; one for \slavemonitor{s} to send the system call states of
their supervised variants to \mastermonitor{}, and one for \mastermonitor{} to
instruct \slavemonitor{s} on how to proceed (abort or continue execution of the
variant).

We developed a technique which we call \emph{asynchronous cross-checking} to
reduce this overhead. Inspired by previous work~\cite{volckaert2016secure,
  koning2016secure}, the idea is to classify system calls into three categories
--- highly sensitive, moderately sensitive, and non-sensitive --- based on the
system call number and/or arguments. With asynchronous cross-checking, highly
sensitive system calls still execute in lock-step, as before. When
\slavemonitor{} deems a system call moderately sensitive, however, it still
sends the system call state information to \mastermonitor{}, but then
immediately resumes execution of the supervised variant without waiting for a
reply from \mastermonitor{}. \mastermonitor{} eventually receives the state
information and may detect a divergence. In that case, \mastermonitor{} will
instruct \slavemonitor{s} to abort execution through a separate error channel
that is used only for this specific purpose. Non-sensitive system calls can
execute without any cross-checking at all.

Note that this policy differs from the selective cross-checking described in
previous work as \toolname{} detects \emph{all} divergences for sensitive system
calls including moderately sensitive ones, whereas ReMon and MvArmor only detect
divergences on invocations of highly sensitive system
calls~\cite{volckaert2016secure, koning2016secure}. There may, however, be a
delay in the detection of divergent invocations of moderately sensitive system
calls.

\paragraph*{\textbf{Immutable State Caching.}} Many system calls, including
\texttt{sys\_getpid} and \texttt{sys\_getppid}, read immutable state such as the
process ID or thread ID. To avoid unnecessary cross-checking and replication,
\toolname{} caches the results of the first invocation of these system
calls. \toolname{} immediately cancels any subsequent invocations of these calls
and returns the cached result instead. Note that, contrary to our other
optimizations, immutable state caching can not be disabled.

\subsection{Reconciling Expected Divergences}
\label{sec:divergences}

In previous NVX systems, all program variants were compiled for the same target
architecture and executed on a single machine. \toolname{}, by contrast, allows
settings where variants target different ISAs/ABIs, and run on different
physical machines. The target ISA and ABI both affect a program's behavior as
observed from the system call interface. ISA/ABI-heterogeneity therefore
challenges the core assumption that variants will behave identically when
provided with identical inputs, as long as the inputs do not trigger a program
bug nor is the program being attacked.

We studied the ABI specification documents and analyzed actual system call
traces of both trivial and complex applications running on ARMv7, ARMv8, i386,
and x86-64 CPUs to understand and anticipate benign behavioral divergences
arising from ISA/ABI-heterogeneity. We incorporated our findings into
\toolname{}'s design and summarize the different classes of expected divergences
and their causes in Table~\ref{tab:divergences}.

\begin{table}[t]
  \centering
  \includegraphics[width=\columnwidth]{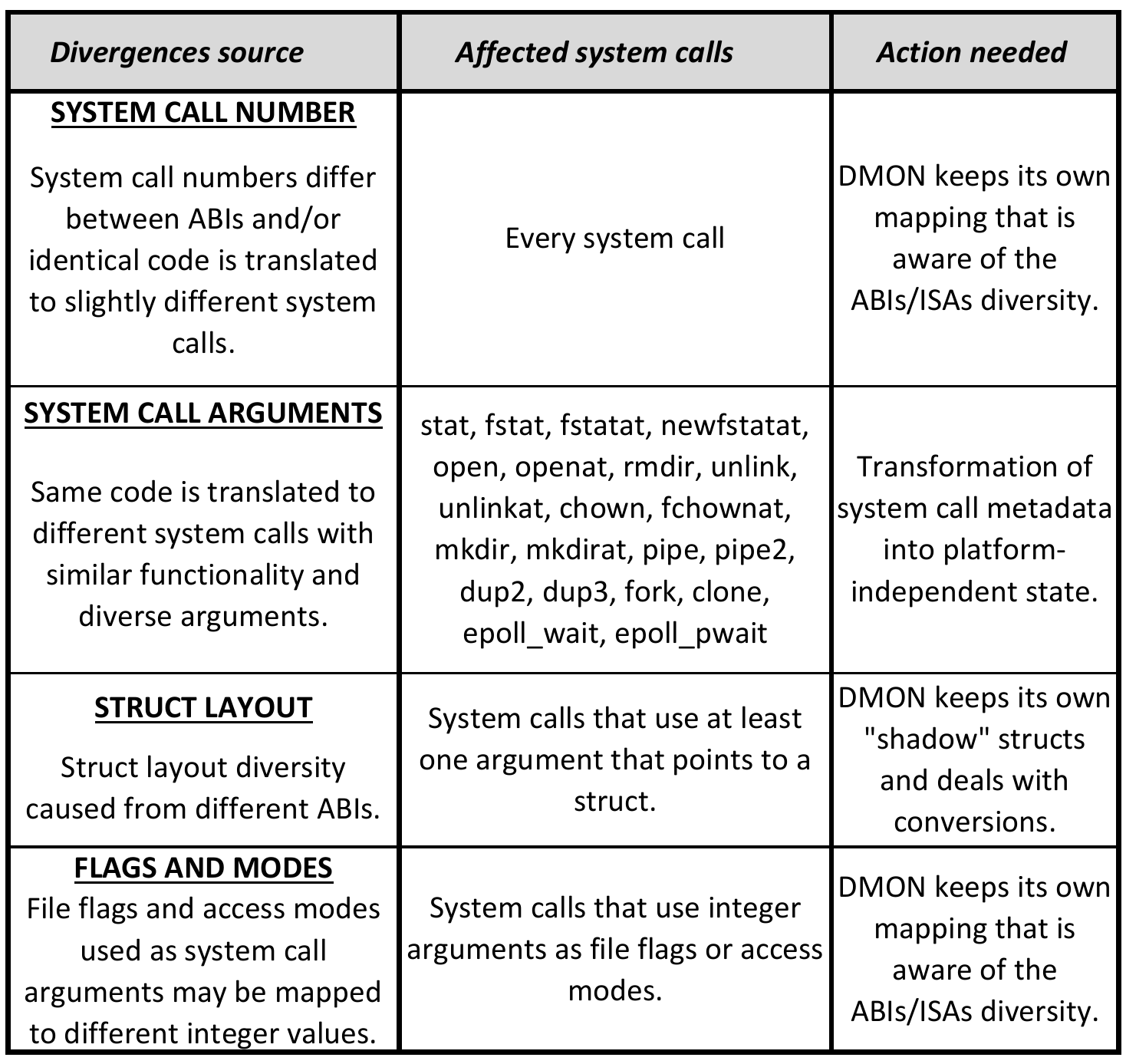}
  \vspace{-1.5em}
  \caption{Categories of expected divergences}
  \label{tab:divergences}
\end{table}

The first and foremost difference between the different ABIs is that the system
call numbers can (and often do) differ between ABIs. The \texttt{sys\_read}
system call, for example, has system call number $0$ on x86-64 platforms and on
ARMv7 platforms that implement the old ARMv7 ABI (i.e., \texttt{arm-linux-gnu}),
$3$ on i386 platforms, and $0x900003$ on ARMv7 platforms that implement the new
ARMv7 ABI (i.e., \texttt{arm-linux-gnueabi}). \toolname{} is aware of these
differences and maps system call numbers to an internal platform-independent
identifier before comparing system call states.



System call arguments are not necessarily bit-for-bit identical across ABIs
either, even when the arguments are in fact the same. This is particularly true
for C structs which may be packed differently (i.e., there may be different
numbers of padding bytes between the struct fields) or have different sizes
depending on the ISA and ABI. To allow for bit-by-bit comparisons of such
structs, \toolname{} converts them to an internal ``shadow'' type that is
carefully specified and/or annotated so it has the same layout on all platforms.

Except for trivial differences in system call numbers and arguments,
heterogeneous-ISA variants may use different system calls altogether, because
not all system calls are available on all platforms. ARMv8 kernels, for example,
do not implement the \texttt{sys\_open} system call. ARMv8 variants therefore
always use \texttt{sys\_openat} to open a file. \texttt{sys\_openat} is similar
to \texttt{sys\_open}, but does have an additional argument that can hold the
file descriptor of a directory. If the \texttt{pathname} argument of the
\texttt{sys\_openat} is relative, then it is interpreted relative to the
directory specified in the additional argument.

x86-64 kernels, on the other hand, implement both \texttt{sys\_open} and
\texttt{sys\_openat} and x86-64 variants regularly use both APIs. Consequently,
we may see divergences in a setup where an ARMv8 and x86-64 variant try to open
the same files. \toolname{} deals with these divergence by transforming the
system call states for similar system calls into generic platform-independent
states, prior to cross-checking. In this concrete example, \toolname{} would
fully resolve the paths the variants are trying to access and then build system
call states corresponding with the invocation of a \texttt{sys\_openat} call,
regardless of whether the variant called \texttt{sys\_open} or
\texttt{sys\_openat}.

Finally, there are several system calls that accept flags as their
arguments. These flags can specify file access modes, file statuses, etc. The
integer values assigned to these flags may differ across ABIs. Consequently,
cross-checking these flag values may trigger false positive detections. Once
again, \toolname{} deals with these divergences by mapping all flag values to an
internal platform-independent value prior to cross-checking.

\section{\toolname{} Implementation}
\label{sec:implementation}


We implemented \toolname{} for GNU/Linux. \toolname{} runs natively on the
x86-64 and ARMv8 architectures and supports variants compiled for these
architectures. \toolname{} also has partial support for ARMv7 and i386.

\paragraph*{\textbf{System Call Support and Classification.}} \toolname{}
currently has cross-checking and replication support for \numofsupsyscalls{}
system calls. Table~\ref{tab:syscalls} shows an overview of these calls, and how
\toolname{} cross-checks and replicates them. The type of cross-checking depends
on the security-sensitivity of the call (see
Section~\ref{sec:optimizations}). \toolname{} always cross-checks highly
sensitive system calls in lock-step. Moderately sensitive calls are either
checked asynchronously, if the asynchronous cross-checking optimization is
enabled, or in lock-step if the optimization is disabled. None-sensitive calls
are not checked at all.

The type of replication depends on the kind of results the system call returns.
\toolname{} enforces replication for all I/O operations that are not reads from
static files (see Section~\ref{sec:optimizations}), and for all system calls
that return mutable program state. Read operations from static files execute
without replication if the permissive filesystem access optimization is enabled.
System calls that must be executed by all variants and system calls that read
cached immutable program state are not subject to any replication.

\begin{table}[t]
  \centering
  \includegraphics[width=\columnwidth]{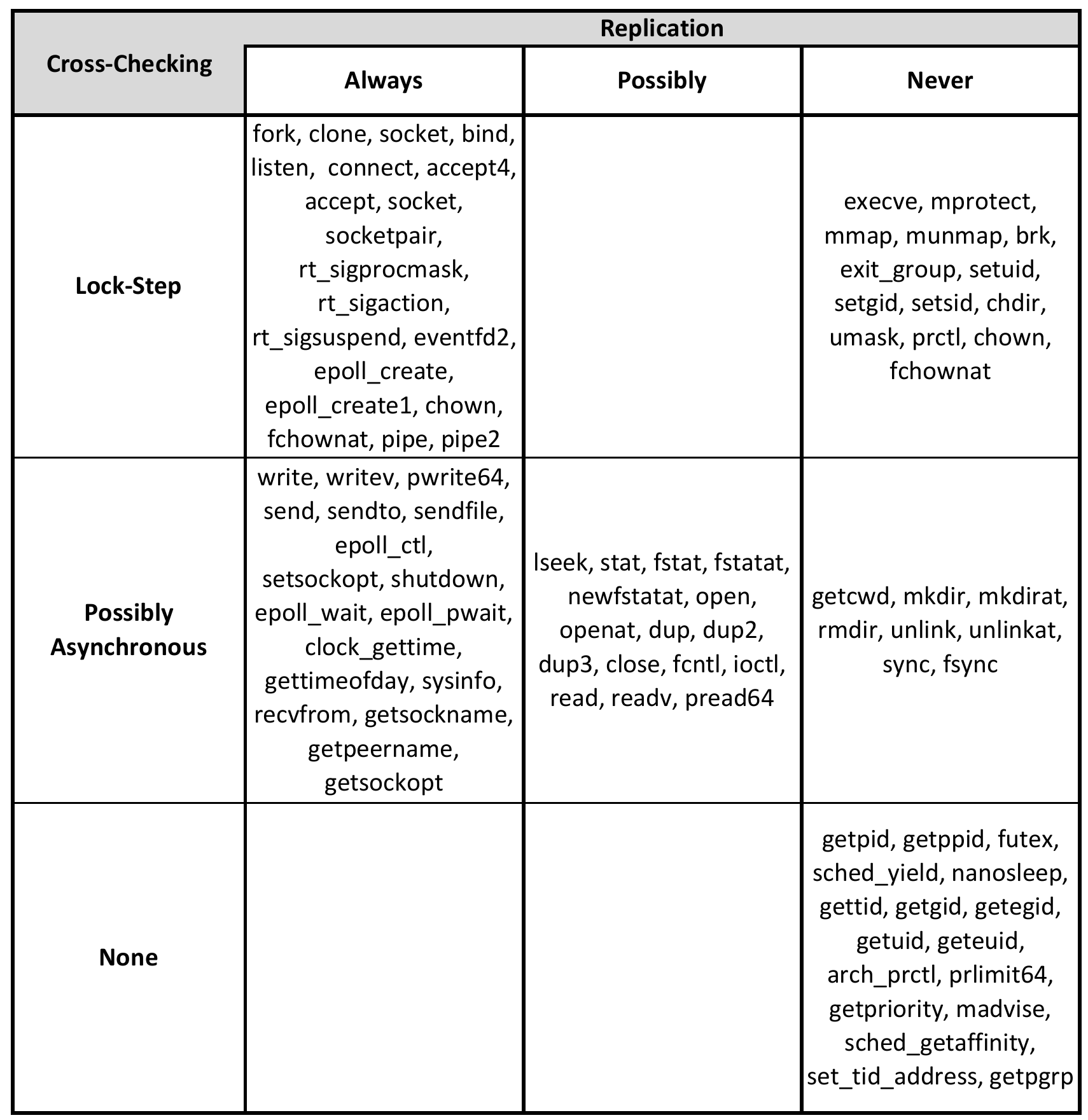}
  \vspace{-1.5em}
  \caption{Supported system calls and their classification}
  \label{tab:syscalls}
\end{table}

\paragraph*{\textbf{Support Infrastructure.}} Aside from the core components
described in the previous section, \toolname{} comes with two support
components. \daemon{} is a standalone service that needs to be installed as a
Unix daemon. \daemon{} acts on service requests from our second component, the
\appstarter{}. Depending on the type of service request, \daemon{} either
launches a new monitor and one or more variants, or shuts down a running monitor
and its variants. The \appstarter{} component reads a configuration file written
by the system administrator and then sends the appropriate service requests to
the \daemon{} components. The configuration file contains information on which
variants to start or shut down (e.g., their architectures, path names, command
line arguments, etc.), and which machines the variants should run on.

\paragraph*{\textbf{Virtual System Calls.}} On most architectures, Linux loads
a \textbf{Virtual Dynamic Shared Object (VDSO)} or \textbf{vsyscall} page into
the address spaces of all user-space programs. These executable code pages
expose a number of so-called virtual system calls, which allow the program to
execute certain system calls (e.g., \texttt{sys\_gettimeofday}) without mode
switching into kernel space. Most NVX systems either hide, replace, or disable
the VDSO and vsyscall page because virtual system calls are invisible to the
monitor, and therefore bypass the replication mechanism. For our prototype, we
patched the C library our variants link against so that the library never uses
virtual system calls. Operations such as \texttt{gettimeofday} are therefore
always visible to our monitors.

\section{Limitations}


\paragraph*{\textbf{Asynchronous Signals.}} Asynchronous signals are currently
not supported in our implementation. Our monitors discard any asynchronous signals that the variants might
receive. This not a fundamental limitation of our approach but a matter of additional engineering effort.
Signal handling mechanisms would be implemented as has already been
described in earlier work~\cite{salamat2009orchestra, hosek2015varan}.

\paragraph*{\textbf{Multithreading.}} \toolname{} has only partial support
for multithreaded variants. Running multithreaded variants currently may trigger
divergences and false positive detections. Earlier work describes techniques to
support variants in which the threads do not communicate
directly~\cite{hosek2015varan,koning2016secure,xu2017bunshin,lu2018stopping}. These
techniques can directly be incorporated into our prototype. Supporting variants in
which the threads \emph{do} communicate directly is more difficult. The current
state-of-the-art is to capture the order in which the leader variant executes
thread synchronization instructions, and to replay this order in the follower
variants~\cite{volckaert2017taming}. Incorporating this mechanism into our
prototype would require substantial engineering effort.

\paragraph*{\textbf{Address-Dependent Behavior.}} Similarly, there might be
divergences in programs featuring address-dependent behavior. If a program
inserts objects into a binary tree based on their memory addresses, and
subsequently prints out that tree, the output would likely differ across
variants. Earlier work describes several similar
cases~\cite{volckaert2012ghumvee}. We are not aware of any automated solutions
for identifying and neutralizing such address-dependent computations, so this is
currently an open problem that applies to all NVX systems.


\section{Security Analysis}

\paragraph*{Scope.}

NVX systems can prevent the attacker from using \emph{absolute} code addresses
in an exploit payload by adopting a technique called Address Space Partitioning
(ASP)~\cite{cox2006n, volckaert2016cloning, lu2018stopping}. With ASP, the NVX
system lays out the variants' address spaces in such a way that (i) the base
addresses of executable code regions are randomly chosen, and (ii) no absolute
memory address ever points to valid code in more than one variant. We refer the
reader to earlier work for a detailed security analysis of this
technique~\cite{cox2006n, volckaert2016cloning, lu2018stopping}, and focus on
evaluating the additional security \toolname{} can provide through
ISA/ABI-heterogeneity. Specifically, we show the extent to which
ISA/ABI-heterogeneity prevents concrete code-reuse and data-only attacks that
cannot be easily stopped using existing NVX systems.

\paragraph*{Analysis Targets and Configurations.}

We used four popular server applications --- Nginx 1.14.2, Lighttpd 1.4.52, Redis
5.0.1, and ProFTPD 1.3.0 --- as our analysis targets, 
which is in line with previous work on security-oriented NVX
systems~\cite{volckaert2016secure,koning2016secure,xu2017bunshin,lu2018stopping}.
We evaluated the security of a heterogeneous configuration with one program
variant compiled for Intel x86-64 and one for ARMv7.

\subsection{Code Layout Diversity}
\label{sec:security:code}

Existing NVX systems that deploy ASP can be bypassed using attacks that rely on
partial overwrites of code pointers such as return addresses or function
pointers~\cite{phrack2002bypassing, goktas2018pirop}. The basic idea is to force
the program to produce a (number of) legal code pointer(s) at memory locations
that the attacker can overwrite. The attacker then overwrites the least
significant bits or adds arbitrary offsets to each of these code pointers, and
thereby diverts the execution of the program to a series of attacker-chosen
gadgets (i.e., instruction sequences ending with indirect branches, such as
return instructions). In the PIROP attack, for example, Goktas et al. exploited
a vulnerability in the Asterisk communication server that allowed them to
produce legal return addresses at an attacker-controlled position on the
stack~\cite{goktas2018pirop}. They then overwrote the least significant byte of
each of these return addresses to build a so-called PIROP gadget chain, which
they then invoked by exploiting another vulnerability.

The reason why these attacks bypass existing NVX systems is because they do not
require any information leakage (which the NVX system would detect), and because
the same partial pointer overwrites can achieve the same results in each
variant. In this section, we show that \toolname{} makes these
position-independent code-reuse attacks far more challenging because
ISA/ABI-heterogeneity substantially reduces the number of position-independent
gadgets available to the attacker.

\paragraph*{Position-Independent Gadget Availability.}

Position-independent gadgets are instruction sequences that can be
\emph{reliably} invoked by patching legal code pointers. We consider two ways to
patch legal code pointers. First, an attacker could overwrite an offset variable
that is later added to a code pointer in a pointer arithmetic operation. This
primitive allows attackers to reliably invoke any gadget, as long as the
internal layout of the target binary is known.

Second, the attacker could overwrite the least significant bits of a code
pointer directly using a memory write vulnerability. This primitive is far less
potent than the former, as it allows the attacker to overwrite only the $8$
least significant bits (i.e., one byte). Overwriting more than one byte is not
possible unless the attacker knows the base address of the target binary because
the ASP scheme randomizes all but the $12$ least significant bits of each base
address.

We compiled a list of the position-independent gadgets in both our x86-64 and
ARMv7 binaries as follows. We first collected the addresses of (i) all
instructions that immediately follow call instructions, and (ii) all
address-taken functions in the program. The former is an approximation of the
set of legal return addresses that could exist in the program's address space at
any given point during its execution. The latter is the set of other code
pointers that could be found in the program's memory. Combined, this list
approximates the set of pointers that \emph{could} potentially be patched by
attackers to construct position-independent code-reuse payloads.

We then used Ropper to generate lists of regular ROP gadgets consisting of $15$
instructions or less~\cite{ropper}. This, again, is consistent with related
work~\cite{goktas2018pirop}.


Next, we combined the two lists for each binary as follows. For every code
pointer in the first list, we calculated the (i) addresses of all gadgets
relative to the pointer, and (ii) absolute addresses of gadgets that only differ
from the code pointer in their $8$ least significant bits. The former is the set
of gadgets reachable through offset overwrites, while the latter is the set of
gadgets reachable through partial pointer overwrites.

Next we correlated the position-independent gadgets found for the x86-64 binary
with those found for ARMv7. For each x86-64 gadget, we checked whether there is
an ARMv7 gadget that can be reached using the same offset overwrite/partial
pointer overwrite. We then eliminated gadgets whose absolute address or offset
from the source code pointer is not $4$-byte aligned, since code pointers
patched in either way would be unaligned on ARMv7 and would trigger an unaligned
instruction exception when the gadget is invoked.

\pgfplotsset{
  every tick label/.append style={font=\scriptsize},
  every axis/.append style={font=\footnotesize},
}
\pgfplotstableread[col sep=comma]{pirop-gadgets.csv}{\numgadgets}

\pgfplotstablegetelem{0}{num-base-ptrs}\of{\numgadgets}
\pgfmathsetmacro{\nginxnumptrs}{\pgfplotsretval}
\pgfplotstablegetelem{1}{num-base-ptrs}\of{\numgadgets}
\pgfmathsetmacro{\lighttpdnumptrs}{\pgfplotsretval}
\pgfplotstablegetelem{2}{num-base-ptrs}\of{\numgadgets}
\pgfmathsetmacro{\redisnumptrs}{\pgfplotsretval}
\pgfplotstablegetelem{3}{num-base-ptrs}\of{\numgadgets}
\pgfmathsetmacro{\proftpdnumptrs}{\pgfplotsretval}

\pgfplotstablegetelem{0}{off-wr-surviving-mean}\of{\numgadgets}
\pgfmathsetmacro{\nginxoffwr}{\pgfplotsretval}
\pgfplotstablegetelem{0}{off-wr-reduction-mean}\of{\numgadgets}
\pgfmathsetmacro{\nginxoffwrreduction}{\pgfplotsretval}
\pgfplotstablegetelem{0}{off-wr-mean-ratio}\of{\numgadgets}
\pgfmathsetmacro{\nginxoffwrratio}{\pgfplotsretval}
\pgfplotstablegetelem{1}{off-wr-surviving-mean}\of{\numgadgets}
\pgfmathsetmacro{\lighttpdoffwr}{\pgfplotsretval}
\pgfplotstablegetelem{1}{off-wr-reduction-mean}\of{\numgadgets}
\pgfmathsetmacro{\lighttpdoffwrreduction}{\pgfplotsretval}
\pgfplotstablegetelem{1}{off-wr-mean-ratio}\of{\numgadgets}
\pgfmathsetmacro{\lighttpdoffwrratio}{\pgfplotsretval}
\pgfplotstablegetelem{2}{off-wr-surviving-mean}\of{\numgadgets}
\pgfmathsetmacro{\redisoffwr}{\pgfplotsretval}
\pgfplotstablegetelem{2}{off-wr-reduction-mean}\of{\numgadgets}
\pgfmathsetmacro{\redisoffwrreduction}{\pgfplotsretval}
\pgfplotstablegetelem{2}{off-wr-mean-ratio}\of{\numgadgets}
\pgfmathsetmacro{\redisoffwrratio}{\pgfplotsretval}
\pgfplotstablegetelem{3}{off-wr-surviving-mean}\of{\numgadgets}
\pgfmathsetmacro{\proftpdoffwr}{\pgfplotsretval}
\pgfplotstablegetelem{3}{off-wr-reduction-mean}\of{\numgadgets}
\pgfmathsetmacro{\proftpdoffwrreduction}{\pgfplotsretval}
\pgfplotstablegetelem{3}{off-wr-mean-ratio}\of{\numgadgets}
\pgfmathsetmacro{\proftpdoffwrratio}{\pgfplotsretval}

\pgfplotstablegetelem{0}{ptr-wr-surviving-mean}\of{\numgadgets}
\pgfmathsetmacro{\nginxptrwr}{\pgfplotsretval}
\pgfplotstablegetelem{0}{ptr-wr-reduction-mean}\of{\numgadgets}
\pgfmathsetmacro{\nginxptrwrreduction}{\pgfplotsretval}
\pgfplotstablegetelem{0}{ptr-wr-mean-ratio}\of{\numgadgets}
\pgfmathsetmacro{\nginxptrwrratio}{\pgfplotsretval}
\pgfplotstablegetelem{1}{ptr-wr-surviving-mean}\of{\numgadgets}
\pgfmathsetmacro{\lighttpdptrwr}{\pgfplotsretval}
\pgfplotstablegetelem{1}{ptr-wr-reduction-mean}\of{\numgadgets}
\pgfmathsetmacro{\lighttpdptrwrreduction}{\pgfplotsretval}
\pgfplotstablegetelem{1}{ptr-wr-mean-ratio}\of{\numgadgets}
\pgfmathsetmacro{\lighttpdptrwrratio}{\pgfplotsretval}
\pgfplotstablegetelem{2}{ptr-wr-surviving-mean}\of{\numgadgets}
\pgfmathsetmacro{\redisptrwr}{\pgfplotsretval}
\pgfplotstablegetelem{2}{ptr-wr-reduction-mean}\of{\numgadgets}
\pgfmathsetmacro{\redisptrwrreduction}{\pgfplotsretval}
\pgfplotstablegetelem{2}{ptr-wr-mean-ratio}\of{\numgadgets}
\pgfmathsetmacro{\redisptrwrratio}{\pgfplotsretval}
\pgfplotstablegetelem{3}{ptr-wr-surviving-mean}\of{\numgadgets}
\pgfmathsetmacro{\proftpdptrwr}{\pgfplotsretval}
\pgfplotstablegetelem{3}{ptr-wr-reduction-mean}\of{\numgadgets}
\pgfmathsetmacro{\proftpdptrwrreduction}{\pgfplotsretval}
\pgfplotstablegetelem{3}{ptr-wr-mean-ratio}\of{\numgadgets}
\pgfmathsetmacro{\proftpdptrwrratio}{\pgfplotsretval}

\pgfplotstableread[col sep=comma]{pirop-gadgets-syscallnum.csv}{\numgadgetssyscallnum}
\pgfplotstableread[col sep=comma]{pirop-gadgets-arg1.csv}{\numgadgetsfirstarg}
\pgfplotstableread[col sep=comma]{pirop-gadgets-arg2.csv}{\numgadgetssecondarg}
\pgfplotstableread[col sep=comma]{pirop-gadgets-arg3.csv}{\numgadgetsthirdarg}

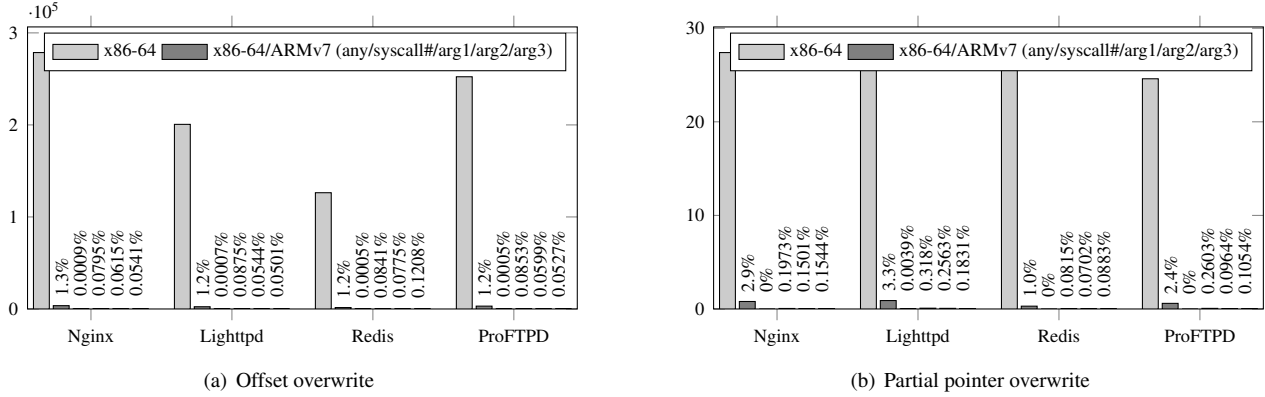
\begin{figure*}[htbp]
  \centering
  \subfigure[Offset overwrite]{
  \begin{tikzpicture}
    \begin{axis}[
      ybar=1.5pt,
      bar width=6pt,
      enlarge x limits=0.15,
      area legend,
      height=0.3\textwidth,
      width=0.5\textwidth,
      legend columns=-1,
      legend style={font=\scriptsize},
      symbolic x coords={
        Nginx,
        Lighttpd,
        Redis,
        ProFTPD
      },
      xtick=data,
      ymin=0
      ]
      \addplot [
        fill=white!80!black,
      ] plot [
      ] table [
        skip first n=1,
        x index=0,
        y index=2,
      ]{\numgadgets};
      \addplot [
        fill=white!50!black,
      ] plot [
      ] table [
        skip first n=1,
        x index=0,
        y index=5,
      ]{\numgadgets};
      \addplot [
        fill=white!70!black,
      ] plot [
      ] table [
        skip first n=1,
        x index=0,
        y index=6,
      ]{\numgadgetssyscallnum};
      \addplot [
        fill=white!80!black,
      ] plot [
      ] table [
        skip first n=1,
        x index=0,
        y index=6,
      ]{\numgadgetsfirstarg};
      \addplot [
        fill=white!90!black,
      ] plot [
      ] table [
        skip first n=1,
        x index=0,
        y index=6,
      ]{\numgadgetssecondarg};
      \addplot [
        fill=white!100!black,
      ] plot [
      ] table [
        skip first n=1,
        x index=0,
        y index=6,
      ]{\numgadgetsthirdarg};
      \node[
        xshift=-11pt,
        black, above, style={font=\scriptsize}
      ] at (axis cs: Nginx,\nginxoffwr)
           {\rotatebox{90}{\nginxoffwrratio\%}};
      \node[
        xshift=-4pt,
        black, above, style={font=\scriptsize}
      ] at (axis cs: Nginx,\nginxoffwr)
           {\rotatebox{90}{0.0009\%}};
      \node[
        xshift=3pt,
        black, above, style={font=\scriptsize}
      ] at (axis cs: Nginx,\nginxoffwr)
           {\rotatebox{90}{0.0795\%}};
      \node[
        xshift=10pt,
        black, above, style={font=\scriptsize}
      ] at (axis cs: Nginx,\nginxoffwr)
           {\rotatebox{90}{0.0615\%}};
      \node[
        xshift=17pt,
        black, above, style={font=\scriptsize}
      ] at (axis cs: Nginx,\nginxoffwr)
           {\rotatebox{90}{0.0541\%}};
      \node[
        xshift=-11pt,
        black, above, style={font=\scriptsize}
      ] at (axis cs: Lighttpd,\lighttpdoffwr)
           {\rotatebox{90}{\lighttpdoffwrratio\%}};
      \node[
        xshift=-4pt,
        black, above, style={font=\scriptsize}
      ] at (axis cs: Lighttpd,\lighttpdoffwr)
           {\rotatebox{90}{0.0007\%}};
      \node[
        xshift=3pt,
        black, above, style={font=\scriptsize}
      ] at (axis cs: Lighttpd,\lighttpdoffwr)
           {\rotatebox{90}{0.0875\%}};
      \node[
        xshift=10pt,
        black, above, style={font=\scriptsize}
      ] at (axis cs: Lighttpd,\lighttpdoffwr)
           {\rotatebox{90}{0.0544\%}};
      \node[
        xshift=17pt,
        black, above, style={font=\scriptsize}
      ] at (axis cs: Lighttpd,\lighttpdoffwr)
           {\rotatebox{90}{0.0501\%}};
      \node[
        xshift=-11pt,
        black, above, style={font=\scriptsize}
      ] at (axis cs: Redis,\redisoffwr)
           {\rotatebox{90}{\redisoffwrratio\%}};
      \node[
        xshift=-4pt,
        black, above, style={font=\scriptsize}
      ] at (axis cs: Redis,\redisoffwr)
           {\rotatebox{90}{0.0005\%}};
      \node[
        xshift=3pt,
        black, above, style={font=\scriptsize}
      ] at (axis cs: Redis,\redisoffwr)
           {\rotatebox{90}{0.0841\%}};
      \node[
        xshift=10pt,
        black, above, style={font=\scriptsize}
      ] at (axis cs: Redis,\redisoffwr)
           {\rotatebox{90}{0.0775\%}};
      \node[
        xshift=17pt,
        black, above, style={font=\scriptsize}
      ] at (axis cs: Redis,\redisoffwr)
           {\rotatebox{90}{0.1208\%}};
      \node[
        xshift=-11pt,
        black, above, style={font=\scriptsize}
      ] at (axis cs: ProFTPD,\proftpdoffwr)
           {\rotatebox{90}{\proftpdoffwrratio\%}};
      \node[
        xshift=-4pt,
        black, above, style={font=\scriptsize}
      ] at (axis cs: ProFTPD,\proftpdoffwr)
           {\rotatebox{90}{0.0005\%}};
      \node[
        xshift=3pt,
        black, above, style={font=\scriptsize}
      ] at (axis cs: ProFTPD,\proftpdoffwr)
           {\rotatebox{90}{0.0853\%}};
      \node[
        xshift=10pt,
        black, above, style={font=\scriptsize}
      ] at (axis cs: ProFTPD,\proftpdoffwr)
           {\rotatebox{90}{0.0599\%}};
      \node[
        xshift=17pt,
        black, above, style={font=\scriptsize}
      ] at (axis cs: ProFTPD,\proftpdoffwr)
           {\rotatebox{90}{0.0527\%}};
      \legend{
        x86-64,
        x86-64/ARMv7 (any/syscall\#/arg1/arg2/arg3)
      }
    \end{axis}
  \end{tikzpicture}
  }\hspace{3em}
  \subfigure[Partial pointer overwrite]{
  \begin{tikzpicture}
    \begin{axis}[
      ybar=1.5pt,
      bar width=6pt,
      enlarge x limits=0.15,
      area legend,
      height=0.3\textwidth,
      width=0.5\textwidth,
      legend columns=-1,
      legend style={font=\scriptsize},
      symbolic x coords={
        Nginx,
        Lighttpd,
        Redis,
        ProFTPD
      },
      xtick=data,
      ymin=0
      ]
      \addplot [
        fill=white!80!black,
      ] plot [
      ] table [
        skip first n=1,
        x index=0,
        y index=14,
      ]{\numgadgets};
      \addplot [
        fill=white!50!black,
      ] plot [
      ] table [
        skip first n=1,
        x index=0,
        y index=17,
      ]{\numgadgets};
      \addplot [
        fill=white!70!black,
      ] plot [
      ] table [
        skip first n=1,
        x index=0,
        y index=13,
      ]{\numgadgetssyscallnum};
      \addplot [
        fill=white!80!black,
      ] plot [
      ] table [
        skip first n=1,
        x index=0,
        y index=13,
      ]{\numgadgetsfirstarg};
      \addplot [
        fill=white!90!black,
      ] plot [
      ] table [
        skip first n=1,
        x index=0,
        y index=13,
      ]{\numgadgetssecondarg};
      \addplot [
        fill=white!100!black,
      ] plot [
      ] table [
        skip first n=1,
        x index=0,
        y index=13,
      ]{\numgadgetsthirdarg};
      \node[
        xshift=-11pt,
        black, above, style={font=\scriptsize}
      ] at (axis cs: Nginx,\nginxptrwr)
           {\rotatebox{90}{\nginxptrwrratio\%}};
      \node[
        xshift=-4pt,
        black, above, style={font=\scriptsize}
      ] at (axis cs: Nginx,\nginxptrwr)
           {\rotatebox{90}{0\%}};
      \node[
        xshift=3pt,
        black, above, style={font=\scriptsize}
      ] at (axis cs: Nginx,\nginxptrwr)
           {\rotatebox{90}{0.1973\%}};
      \node[
        xshift=10pt,
        black, above, style={font=\scriptsize}
      ] at (axis cs: Nginx,\nginxptrwr)
           {\rotatebox{90}{0.1501\%}};
      \node[
        xshift=17pt,
        black, above, style={font=\scriptsize}
      ] at (axis cs: Nginx,\nginxptrwr)
           {\rotatebox{90}{0.1544\%}};
      \node[
        xshift=-11pt,
        black, above, style={font=\scriptsize}
      ] at (axis cs: Lighttpd,\lighttpdptrwr)
           {\rotatebox{90}{\lighttpdptrwrratio\%}};
      \node[
        xshift=-4pt,
        black, above, style={font=\scriptsize}
      ] at (axis cs: Lighttpd,\lighttpdptrwr)
           {\rotatebox{90}{0.0039\%}};
      \node[
        xshift=3pt,
        black, above, style={font=\scriptsize}
      ] at (axis cs: Lighttpd,\lighttpdptrwr)
           {\rotatebox{90}{0.318\%}};
      \node[
        xshift=10pt,
        black, above, style={font=\scriptsize}
      ] at (axis cs: Lighttpd,\lighttpdptrwr)
           {\rotatebox{90}{0.2563\%}};
      \node[
        xshift=17pt,
        black, above, style={font=\scriptsize}
      ] at (axis cs: Lighttpd,\lighttpdptrwr)
           {\rotatebox{90}{0.1831\%}};
      \node[
        xshift=-11pt,
        black, above, style={font=\scriptsize}
      ] at (axis cs: Redis,\redisptrwr)
           {\rotatebox{90}{\redisptrwrratio\%}};
      \node[
        xshift=-4pt,
        black, above, style={font=\scriptsize}
      ] at (axis cs: Redis,\redisptrwr)
           {\rotatebox{90}{0\%}};
      \node[
        xshift=3pt,
        black, above, style={font=\scriptsize}
      ] at (axis cs: Redis,\redisptrwr)
           {\rotatebox{90}{0.0815\%}};
      \node[
        xshift=10pt,
        black, above, style={font=\scriptsize}
      ] at (axis cs: Redis,\redisptrwr)
           {\rotatebox{90}{0.0702\%}};
      \node[
        xshift=17pt,
        black, above, style={font=\scriptsize}
      ] at (axis cs: Redis,\redisptrwr)
           {\rotatebox{90}{0.0883\%}};
      \node[
        xshift=-11pt,
        black, above, style={font=\scriptsize}
      ] at (axis cs: ProFTPD,\proftpdptrwr)
           {\rotatebox{90}{\proftpdptrwrratio\%}};
      \node[
        xshift=-4pt,
        black, above, style={font=\scriptsize}
      ] at (axis cs: ProFTPD,\proftpdptrwr)
           {\rotatebox{90}{0\%}};
      \node[
        xshift=3pt,
        black, above, style={font=\scriptsize}
      ] at (axis cs: ProFTPD,\proftpdptrwr)
           {\rotatebox{90}{0.2603\%}};
      \node[
        xshift=10pt,
        black, above, style={font=\scriptsize}
      ] at (axis cs: ProFTPD,\proftpdptrwr)
           {\rotatebox{90}{0.0964\%}};
      \node[
        xshift=17pt,
        black, above, style={font=\scriptsize}
      ] at (axis cs: ProFTPD,\proftpdptrwr)
           {\rotatebox{90}{0.1054\%}};
      \legend{
        x86-64,
        x86-64/ARMv7 (any/syscall\#/arg1/arg2/arg3)
      }
    \end{axis}
  \end{tikzpicture}
  }
  \caption{The average number of position-independent code-reuse gadgets
    available from each code pointer for each pointer patching strategy.}
  \label{fig:pirop-gadget}
  \vspace{-0.3cm}
\end{figure*}

We collected \nginxnumptrs{} code pointers from Nginx, \lighttpdnumptrs{} code
pointers from Lighttpd, \redisnumptrs{} code pointers from Redis, and
\proftpdnumptrs{} code pointers from ProFTPD. Figure~\ref{fig:pirop-gadget}
shows how many gadgets can be reached on average from each code pointer by
offset overwrite and partial pointer overwrite attacks. In a traditional NVX
system where all variants are compiled for Intel x86-64, all of the gadgets
identified in the x86-64 binary would survive. In contrast, in all four of our
target programs, and for both code pointer patching strategies, less than 3.3\%
of the gadgets survive in an NVX configuration with a x86-64 variant and an
ARMv7 variant.

\paragraph*{Position-Independent Gadget Semantics.}

The final step of an exploit is often to call a security-sensitive function or a
system call with attacker-specified arguments (e.g., \code{execve} with the
\code{/bin/sh} string as the first argument to spawn a shell). The
ABI-heterogeneity provided by \toolname{} imposes another constraint on chaining
gadgets to build such an exploit. Because different architectures have different
calling conventions for system calls and subroutines, as shown in
Table~\ref{tab:callingconv} and Table~\ref{tab:syscallconv}, the attacker should
chain a sequence of gadgets that prepare the same set of arguments, but in a
different way for each architecture.

\begin{table}[t]
  \centering
  \caption{Subroutine calling conventions for x86-64 and ARMv7.\vspace{0.5em}}
  \label{tab:callingconv}
  \resizebox{0.48\textwidth}{!}{
  \begin{tabular}{|l|l|l|l|l|l|l|l|l|}
    \hline
arch/ABI &     arg1    &  arg2    &  arg3    & arg4    & arg5    & arg6    & arg7     & result \\\hline
x86-64   &     \code{rdi} &  \code{rsi} &  \code{rdx} & \code{rcx} & \code{r8}   & \code{r9}   &   -      & \code{rax} \\\hline
arm/EABI &     \code{r0}   &  \code{r1}   &  \code{r2}   & \code{r3}   & stack   & stack   & stack    & \code{r0}-\code{r3} \\\hline
  \end{tabular}
  }
\end{table}
\begin{table}[t]
  \centering
  \caption{System call calling conventions for x86-64 and ARMv7.\vspace{0.5em}}
  \label{tab:syscallconv}
  \resizebox{0.48\textwidth}{!}{
  \begin{tabular}{|l|l|l|l|l|l|l|l|l|l|}
    \hline
arch/ABI   & syscall \# &  arg1   &   arg2   &   arg3   &  arg4   &  arg5   &  arg6   &  arg7 & result \\\hline
x86-64     & \code{rax}    & \code{rdi} &  \code{rsi} &  \code{rdx} & \code{r10}  & \code{r8}   & \code{r9}   & - & \code{rax} \\\hline
arm/EABI   & \code{r7}      &  \code{r0}  &   \code{r1}  &  \code{r2}   & \code{r3}   & \code{r4}   & \code{r5}   & \code{r6} & \code{r0} \\\hline
  \end{tabular}
  }
\end{table}
For example, in an ARMv7 variant, the attacker must use \code{r7} to prepare a
system call number, whereas in a x86-64 variant the same attacker must use
\code{rax}. To show the difficulty of constructing a code-reuse attack that
performs one or more system calls and/or subroutine calls, we analyzed the
semantics of position-independent gadgets surviving under
\toolname{}. Specifically, we looked for gadgets that read a value from memory
and write that value into the system call number register, or the registers for
one of the first three arguments of a system or function call. As shown in
Figure~\ref{fig:pirop-gadget}, only a small fraction of the position-independent
gadgets have suitable semantics for
argument preparation (see 3rd to 6th bars in the figure). More interestingly, system call number preparation gadgets
are rare compared to other argument preparation gadgets. In a standalone ARMv7
binary of Nginx, Redis, and ProFTPD, we could not find a single partial-pointer-overwrite based position-independent gadget which
can load a system call number. Obviously then, we also could not find such
gadgets among those that survive across architectures.

\subsection{Data Layout Diversity}




Aside from protecting against code-reuse attacks, \toolname{} also raises the
bar for successful data-only attacks. In a data-only attack, the adversary
forces the program to disclose sensitive information such as pointer values,
corrupt sensitive data such as stored user credentials, or perform arbitrary
computations~\cite{hu2016data, ispoglou2018bop}, without deviating from its
intended control-flow paths.

Prior work showed that NVX systems may be able to stop certain data-only
attacks. First, the system can extend ASP to data regions to prevent the
variants from disclosing absolute pointer values to remote
attackers~\cite{koning2016secure,lu2018stopping}.
Next, the system can run variants with opposing stack growth directions to
thwart attacks that corrupt sensitive stack
variables~\cite{salamat2009orchestra}. Finally, the variants can use randomizing
heap allocators to probabilistically stop data-only attacks that corrupt heap
data~\cite{berger2006diehard}.

All of these defenses provide diversity at page- or object-level granularity.
As such, they cannot prevent intra-object overread or overwrite attacks, in
which a pointer to a specific field in a data structure is used to disclose or
overwrite a different field within that same
structure~\cite{gil2018pointerstretching}. \toolname{}, in
contrast, naturally provides intra-object diversity. Due to differences in sizes
of pointers and primitive data types, as well as differences in struct packing
and alignment, data structures rarely have the same sizes and layouts across
architectures and ABIs.

Previous NVX systems could achieve intra-object diversity by artificially
reorganizing structures at compile time (e.g., by reordering struct fields or
inserting padding between the fields). However, only a limited number of structs
can be compile-time diversified in practice. Specifically, it is not safe to
diversify i) structures used as arguments or return types of external library
functions, ii) structures with an initialization list, iii) structs cast to
different types, etc.~\cite{lin2009polymorphing,chen2016attack}. We examined
struct layouts in a set of Nginx binaries to show how much structure layout
diversity \toolname{} can naturally achieve, compared to the number of
structures that can be compile-time diversified by existing type-based structure
layout randomization techniques~\cite{lin2009polymorphing,chen2016attack}. We
found that existing compile-time techniques could only diversify $50$ out of
$453$ structures, whereas $422$ out of $453$ structures inherently have
different layouts on ARMv7 and x86-64.

\vspace{-.5cm}

\paragraph*{ProFTPD SSL Private Key Leak.}

Hu et al. demonstrated an information disclosure attack on ProFTPD, in which the
attacker locates a base pointer to an SSL context data structure, and then uses
Data-Oriented Programming (DOP) gadgets to traverse through the context and 6
other data structures, ultimately reaching a private key, which is then leaked
to a remote attacker~\cite{hu2016data}. \toolname{} can prevent this attack
because the layouts of the 6 data structures differs across architectures. We
examined the relevant data structures in ARMv7 and x86-64 binaries of ProFTPD
and found that 4 of the 6 pointer fields that need to be dereferenced in this
attack are located at different offsets in the two binaries. A DOP exploit that
traverses through the structs therefore cannot simultaneously reach and leak the
private key on both platforms without triggering an alarm in \toolname{}.

\section{Performance Evaluation}
\label{sec:performanceeval}

We conducted an extensive performance evaluation of \toolname{} using
handwritten microbenchmarks (Section~\ref{sec:microbenchmarks}), as well as
popular high-performance server applications
(Section~\ref{sec:serverbenchmarks}). We ran our benchmarks under two different
configurations:





\paragraph*{The \textbf{low-end configuration}} had an ARMv8 variant running on
a Raspberry Pi 3 Model B board with a quad-core 1.2GHz Broadcom BCM2837 64-bit
CPU and 1GB of RAM, running the 64-bit ARM Debian 9 distribution of GNU/Linux,
as well as an x86-64 variant running on a desktop machine with a quad-core Intel
i5-6500 CPU and 16GB of RAM, running the x86-64 version of Ubuntu 16.04.5
LTS. The machines were connected through a \emph{private} 100 megabit ethernet
connection with approximately 0.5ms latency.

\paragraph*{The \textbf{high-end configuration}} had an x86-64 variant running
on a desktop machine with an octa-core Intel i9-9900K CPU and 32GB of RAM, and
an x86-64 variant running on a machine with a quad-core Intel i5-6500 CPU and
16GB of RAM. Both machines ran the x86-64 version of Ubuntu 16.04.5 LTS and were
connected using a private 100 gigabit connection between two Mellanox ConnectX
ethernet interface cards. These RDMA-capable cards support the Mellanox
Messaging Accelerator, a user-space networking library with TCP support and
sub-microsecond latency.



In both configurations, we ran the leader variant on the slower machine. This
choice minimizes the time the leader variant has to wait in its cross-checking
handlers for the follower variant to send the system call state information
(see Section~\ref{sec:design}).

We evaluated two implementations of our \connector{} component
(see Section~\ref{sec:communication}) for the low-end configuration. The first
implementation, which appears as \textbf{KTCP} (short for kernel-space TCP) in
the graphs, uses the standard Linux TCP/IP stack. The second implementation uses
the \textbf{ENet} protocol. For the high-end configuration, we additionally
evaluated an implementation that leverages the Mellanox Messaging Accelerator
library. This implementation appears as \textbf{UTCP} (short for user-space TCP)
in the graphs. We could not test this UTCP implementation for configuration 1 as
we could not find the appropriate hardware for our ARMv8 board.

We also evaluated the impact of our replication and cross-checking optimizations
described in Section~\ref{sec:optimizations}. Our Asynchronous Cross-Checking
optimization appears as \textbf{ACC} in the graphs, whereas our Permissive
Filesystem Access optimization appears as \textbf{PFA}.

\subsection{Microbenchmarks}
\label{sec:microbenchmarks}


To measure the overhead introduced by \toolname{}, we designed microbenchmarks
to test different combinations of cross-checking and replication strategies (see
Section~\ref{sec:implementation}). The microbenchmarks are small programs that
execute the same system call one million times in a loop.

\begin{figure}[t]
	\centering
	\includegraphics[width=\columnwidth]{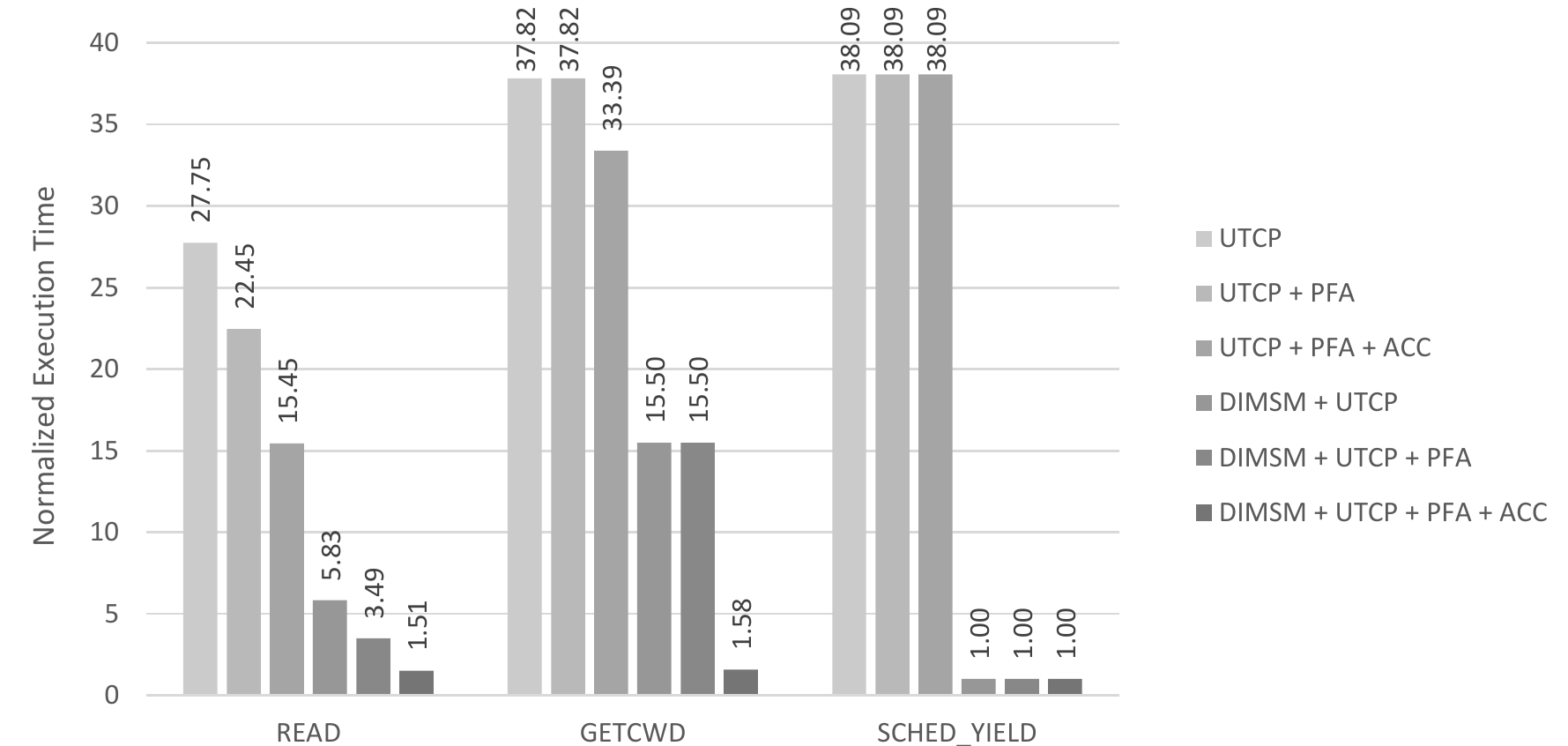}
	\vspace{-.8cm}
	\caption{Microbenchmarks for \emph{high-end} configuration.}
	\label{fig:micro}
	\vspace{-1.50em}
\end{figure}

We used the following system calls:

\begin{enumerate}

	\item \textbf{sys\_read(STATIC\_FILE\_FD, buf, 512)} reads 512 bytes from a
      static file. \toolname{} treats \texttt{sys\_read} as a moderately sensitive
      system call. As such, this microbenchmark benefits from our asynchronous
      cross-checking optimization (see Section~\ref{sec:optimizations}). Since
      the file we are reading is static, \toolname{} also skips replication if
      the permissive file system access optimization is enabled.

	\item \textbf{sys\_getcwd(buf, 512)} retrieves the pathname for the current
      working directory The results of this system call do not need to be
      replicated, as long as the current working directory is either the
      application's root directory, or one of its subdirectories (see
      Section~\ref{sec:optimizations}).
      
	\item \textbf{sys\_sched\_yield()} relinquishes the CPU and moves the
      calling process to the end of the scheduling queue. \toolname{} does not
      perform cross-checking or replication for this system call.

\end{enumerate}

To understand the performance impact of cross-process monitoring in a
distributed setting, we also implemented a rudimentary distributed in-process
monitoring mechanism, similar to the one used in
VARAN~\cite{hosek2015varan}. With this in-process monitoring mechanism, the
variants run in the same address space as their monitors and invoke the
monitor's cross-checking and replication handlers directly, rather than
indirectly through the \texttt{ptrace} API.

For each microbenchmark, we measured the execution time under \toolname{}'s
high-end configuration relative to the native (standalone) execution time
\emph{on the slowest machine}. Figure~\ref{fig:micro} shows the mathematical
average of the run time for three runs of each benchmark, relative to the native
execution time. We used our \textbf{UTCP} implementation of \connector{} for all
experiments, but did run separate tests with and without our permissive file
access (\textbf{PFA}) and asynchronous cross-checking (\textbf{ACC})
optimizations. We then repeated the experiments with the in-process monitoring
mechanism (marked as \textbf{DIMSM} in the graph).

\begin{figure*}[t]
	\centering \includegraphics[width=\linewidth]{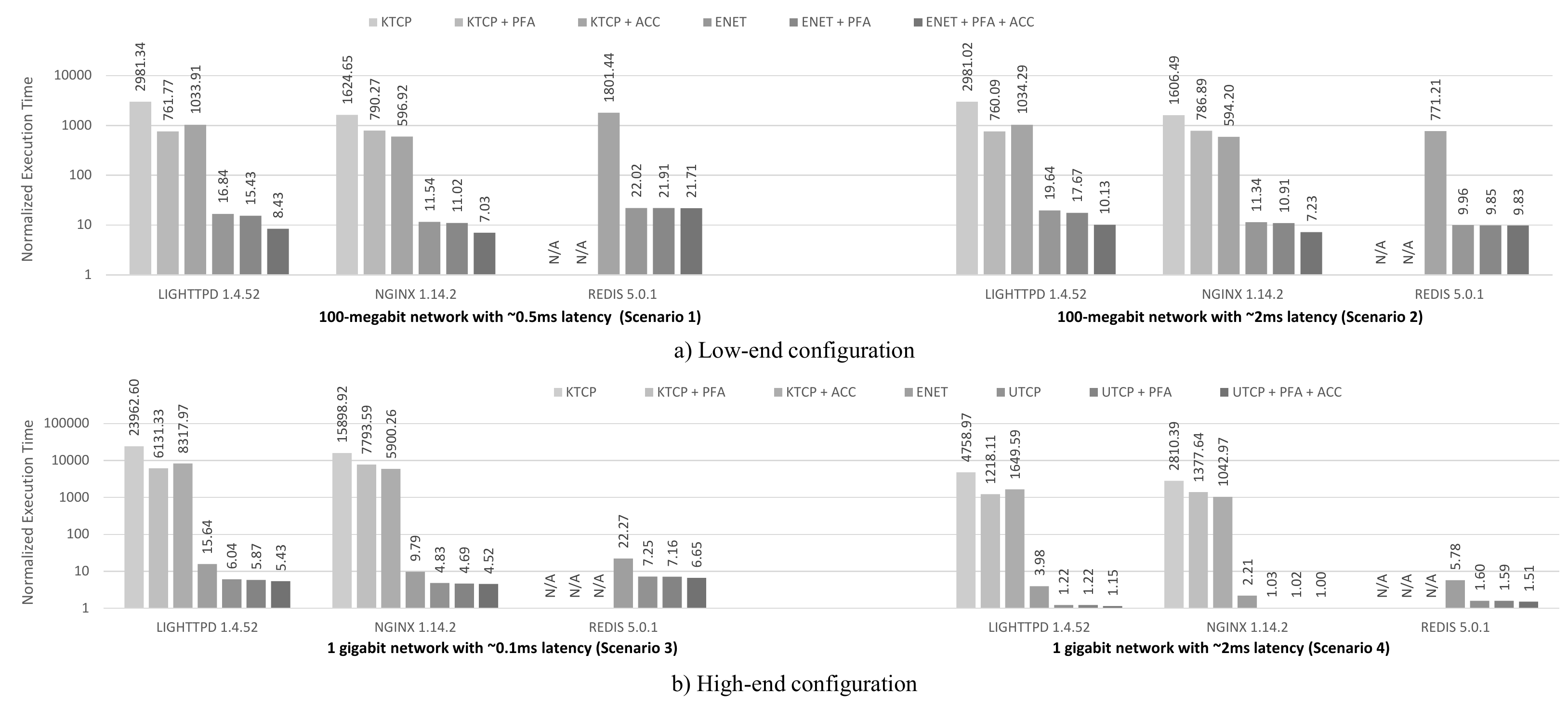}
	\caption{Server benchmarks in two configurations, with two scenarios per configuration.
	}
	\label{fig:servers}
\end{figure*}

The results show that most of the overhead can be attributed to two sources: the
network communication of our replication and cross-checking mechanisms, and the
context switching caused by \texttt{ptrace}. PFA reduces the overhead of our
\texttt{read} benchmark from 28x to 22x, but does not affect the \texttt{getcwd}
and \texttt{sched\_yield} benchmarks. This is unsurprising since \texttt{read}
is the only of the three system calls that accesses static files. ACC further
decreases overhead of \texttt{read} and \texttt{getcwd}, from 22x to 15x and
from 38x to 33x respectively. \texttt{sched\_yield}'s performance is unaffected,
since \toolname{} does not perform any cross-checking for this system call.

The rightmost columns in the graph indicate that the context switching overhead
of \texttt{ptrace} is by far the biggest contributor to \toolname{'s}
overhead. Overhead is reduced from 15x to 1.51x for \texttt{read} and from 33x
to 1.58x for \texttt{getcwd}.  For \texttt{sched\_yield}, there is no observable
overhead.

\subsection{Server Benchmarks}
\label{sec:serverbenchmarks}

We evaluated \toolname{} on 3 popular server applications --- Nginx 1.14.2,
Lighttpd 1.4.52 and Redis 5.0.1 --- that were also used to evaluate prior
work~\cite{hosek2015varan,volckaert2016secure,koning2016secure,lu2018stopping}. For
each of our experiments, we connected a benchmarking client to the machine that
runs the leader variant through a 100 megabit ethernet connection (for our
low-end configuration) or a 1 gigabit ethernet connection (for the high-end
configuration). Figure~\ref{fig:servers} shows our results.

We used the \texttt{wrk} client to measure the throughput and latency for Nginx
and Lighttpd, and the \texttt{redis-benchmark} utility to evaluate Redis. We
configured \texttt{wrk} to repeatedly request the same static 4KB web page for
10 seconds using 10 parallel connections, and \texttt{redis-benchmark} to
simulate 50 clients issuing 100000 requests in total. Completing 100000 requests
takes under a minute when running \texttt{redis} natively on a modern x86
machine. Running the same benchmarks under \toolname{'s} slowest configurations
would take over a day, however, so we decided to not run the benchmarks at all,
and to simply mark the corresponding bars as \texttt{N/A} in the graphs.

We simulated two scenarios for each of our configurations. In \textbf{Scenarios
  1 and 3}, we used the network link between the benchmark client machine and
the machine that runs the leader variant as-is. This meant that the latency on
the 100 megabit link was just under 0.5ms, whereas the latency on the 1 gigabit
link was under 0.1ms.
The benchmarking client was able to fully saturate the leader variant's machine
in both cases. With all of \toolname{'s} optimizations enabled, the performance 
overheads ranged between 7.03x and 21.71x for the low-end configuration, and
between 4.52x and 6.65x for the high-end configuration.

In \textbf{Scenarios 2 and 4}, we used the Linux Traffic Control tool
(\texttt{tc}) and the \texttt{netem} driver to artificially increase the latency
on the benchmarking client's connection to 2ms, thereby simulating the
experimental setups used in prior
work~\cite{cox2006n,volckaert2016secure,maurer2012tachyon}. In these scenarios,
our benchmarking client could not fully saturate the leader variant's
machine. With all of \toolname{'s} optimizations enabled, the performance
overheads ranged between 7.23x and 10.13x for the low-end configuration, and
between 1.0x and 1.51x for the high-end configuration.


We observed two general trends in our results. First, our \connector{}
implementation that uses the standard Linux TCP/IP stack (shown as \textbf{KTCP}
in the graphs) is substantially slower than both of the alternative
implementations. Second, the web server applications benefit greatly from our
permissive filesystem access (\textbf{PFA}) and asynchronous cross-checking
(\textbf{ACC}) optimizations. In scenario 3, for example, PFA decreased the
overhead of Lighttpd from 23692.60x to 6131.33x with the KTCP-based
\connector{}, whereas ACC decreased the overhead from 23692.60x to 8317.97x. For
Redis, the effect of enabling these optimizations is negligible. This makes
sense, because Redis is an in-memory data structure store that rarely accesses
the filesystem.




\subsection{Comparison With Other NVX Systems}

Our server benchmarks show that \toolname{} can incur substantial performance
overheads on saturated machines, even when using the User-Space TCP-based
implementation of our \connector{} component and both of our cross-checking and
replication optimizations. Intuitively, one might think that the inter-monitor
communication overhead is to blame. That is not the case, however, as prior work
showed nearly identical overheads for a traditional single-host NVX
system. Specifically, GHUMVEE was tested on the same server applications (albeit
slightly older versions), and in highly similar circumstances, with a 1 gigabit
link that had less than 0.1ms of latency. GHUMVEE's overhead on Lighttpd was
7.0x on a saturated server (vs 5.43x for \toolname{}), and 12.48x for Redis (vs
6.65x for \toolname{})~\cite{volckaert2016secure}. Handing the monitoring of
non-sensitive system calls over from the \texttt{ptrace}-based GHUMVEE to the
in-process monitor IP-MON brought down the overhead to 2.69x and 1.45x for
Lighttpd and Redis resp. \toolname{} could likely achieve even better
performance if we added full in-process monitoring, because \toolname{} supports
asynchronous cross-checking of moderately sensitive system calls and permissive
filesystem access for \emph{static} content.

\section{Discussion}


\paragraph*{Asymmetrical Attacks.}
%
The heterogeneity present in the code and data layout can still, in theory, be
bypassed by attackers. For example, an attacker can perform a set of malicious
operations, which is functional in one variant, but interpreted as no-operation
in the other. By combining two sets of such operations in an attack (each set
only functional for each variant), the attacker can perform the same malicious
operation in all variants, evading \toolname{}'s detection.

For example, in a PIROP attack, the attacker can find gadgets that are only
functional in one variant, and interpreted as no-operation in the other, and
chain them together. An attacker can also adopt a similar strategy to construct
certain data-only attacks. In a privilege escalation attack, for example, the
attacker can tolerate the difference in the layout of a structure by overwriting
a security sensitive field at both offsets in all variants, even when each
offset is valid only for a single variant. This results in an unintended
corruption of variables in all variants. As long as such corruption does not
trigger an observable divergence, however, the attacker can bypass \toolname{}.



\paragraph*{Leveraging Hardware Features.} A potential advantage of running
variants on different architectures is that the NVX system could leverage
hardware security features available on one platform to protect software running
on other platforms. A future revision of the ARMv8 architecture (ARMv8.5-A), for
example, will include a feature called Memory Tagging. This feature allows the
program to add tags to every pointer and every memory allocation. When the
program dereferences a pointer, the CPU automatically checks if the pointer tag
matches the allocation tag. Memory tagging, if properly implemented, can detect
both spatial and temporal memory errors. A hypothetical configuration in which
\toolname{} runs one variant on an ARMv8.5-A CPU and one variant on an Intel
x86-64 CPU could be used to bring the benefits of memory tagging to Intel x86-64
software. In a similar manner, \toolname{} could enable Intel's Control-flow
Enforcement Technology (CET) for ARM programs, and ARMv8.3's pointer
authentication for Intel x86-64 programs.




\vspace{-0.5em}
\paragraph*{Micro-Architectural Attacks.} While our primary focus was on
defending against memory exploits, we believe \toolname{} might also be able to
stop certain micro-architectural attacks. Rowhammer attacks in particular would
become exceedingly hard to launch against
\toolname{}~\cite{van2016drammer,seaborn2015exploiting,gruss2016rowhammer}. Rowhammer
attacks induce bit flips in so-called weak DRAM cells by rapidly and repeatedly
accessing adjacent DRAM rows. To build reliable Rowhammer attacks, the attacker
needs to know exactly how the memory controller translates physical memory
addresses into DRAM
addresses~\cite{pessl2016drama,tatar2018defeating}. Translation schemes differ
greatly across platforms, however, which makes Rowhammer attack payloads
non-portable. Moreover, even if two machines did have memory controllers using
the same translation scheme, Rowhammer attacks would still fail with high
probability under \toolname{}, as the positions of the weak cells tend to differ
for any given pair of DRAM modules.






\section{Related Work}

\paragraph*{\textbf{N-Variant eXecution.}} The idea of running diversified
software variants in parallel for increased security is not new. Inspired by
Chen and Avizienis' seminal work on N-Version
Programming~\cite{chen1978n,avizienis1985n}, Berger and Zorn proposed a system
for probabilistic memory safety that could run simultaneously execute identical
variants with differently seeded randomizing memory
allocators~\cite{berger2006diehard}. This system only supported applications
that received input through \texttt{stdin} and that wrote output to
\texttt{stdout}, however. Cox et al.'s N-Variant Systems monitored a much wider
array of system calls and replicated I/O from various sources, thus supporting
variants of non-trivial applications such as the Apache web
server~\cite{cox2006n}. Subsequent publications explored consistent delivery of
asynchronous signals~\cite{bruschi2007diversified,salamat2009orchestra}, dealing
with shared memory~\cite{bruschi2007diversified}, thread
synchronization~\cite{volckaert2017taming}, or address-dependent
behavior~\cite{volckaert2012ghumvee}, and new schemes for generating software
variants~\cite{koning2016secure
  ,volckaert2016cloning,xu2017bunshin,lu2018stopping}. Other researchers
suggested to use NVX systems for live patch testing~\cite{maurer2012tachyon,
  hosek2013safe, kim2015dual, hosek2015varan, kwon2016ldx,
  pina2019mvdsua}. Contrary to \toolname{}, however, all of these systems
require that the variants are compiled for the same architecture, and run on the
same host.

Pina et al. proposed a Domain-Specific Language to specify expected divergences
between different variants, and an NVX system that reconciles divergent
variants~\cite{pina2017dsl}. Here, the assumption is that the variants have
minor differences because they were not compiled from the same revision of the
source code. \toolname{} assumes that the variants were compiled from the same
source code, but it does reconcile variants that diverge because of ISA and
ABI-heterogeneity.

\vspace{-1em}
\paragraph*{\textbf{Heterogeneous-ISA Migration.}} Several researchers explored
the idea of heterogeneous-ISA program migration. Devuyst et al. demonstrated
performance and energy efficiency increases by migrating the execution of a
program between the cores of a heterogeneous-ISA
CPU~\cite{devuyst2012execution}. Venkat et al. later showed that
heterogeneous-ISA migration has potential security
benefits~\cite{venkat2016hipstr}. These systems require specialized CPUs that
are not widely available, whereas \toolname{} was designed to run on commodity
hardware.


\vspace{-1em}
\section{Conclusion}

We presented \toolname{}, a novel, distributed N-Variant Execution system that
leverages diversity in instruction set architectures and application binary
interfaces to protect against memory corruption attacks. To bypass \toolname{},
attackers must provide exploits that simultaneously work on two or more
different platforms. We analyzed binaries for two completely different
platforms, one for x86-64 and one for ARMv7 (32-bit). Our analysis shows that
\toolname{} can be an effective mitigation for position-independent code-reuse
attacks which bypass traditional NVX systems. For security-critical
server applications, only 1-3\% of position-independent code-reuse gadgets
survived on both platforms. Furthermore, our case study shows that \toolname{}
can raise the bar for successful data-only attacks when the variants have
different data-layouts due to differences in their data type sizes, struct
packing and alignment rules, etc.


Unlike previous NVX systems which run on a single machine, network communications
between the monitors on different machines become a new source of performance
overhead for \toolname{}. We introduce new optimizations to minimize network
round-trips: permissive filesystem access, asynchronous cross-checking, and
immutable state caching. Our performance evaluation shows that these
optimizations, combined with an optimized network protocol, greatly reduces the
performance overhead (from thousands \textit{times} to single digit
\textit{percentages} in realistic scenarios) without sacrificing \toolname{}'s
security guarantees.



{\normalsize \bibliographystyle{abbrv}
\bibliography{grub}}

\end{document}